# *An Infinite Swapping Approach to the Rare-Event Sampling Problem*

Nuria Plattner and J. D. Doll
Department of Chemistry
Brown University
Providence, RI 02912

Paul Dupuis, Hui Wang and Yufei Liu
Division of Applied Mathematics
Brown University
Providence, RI 02912

J. E. Gubernatis
Theoretical Division
Los Alamos National Laboratory
Los Alamos, NM 87545

*Abstract*

We describe a new approach to the rare-event Monte Carlo sampling problem. This technique utilizes a symmetrization strategy to create probability distributions that are more highly connected and thus more easily sampled than their original, potentially sparse counterparts. After discussing the formal outline of the approach and devising techniques for its practical implementation, we illustrate the utility of the technique with a series of numerical applications to Lennard-Jones clusters of varying complexity and rare-event character.

*I. Introduction:* Monte Carlo methods are among the more versatile and widely utilized tools in modern simulation.[1-5] In particular, they provide a refinable means for treating problems of a physically realistic complexity. They are of importance in classical and quantum statistical-mechanical applications, for example, where the problem of computing thermodynamic properties can be reduced to that of performing well-defined, large-dimensional averages over known probability distributions. Such applications are the primary focus of the present discussion.

A key step in the use of Monte Carlo methods in either a classical or quantum equilibrium statistical-mechanical context is generating a suitable sampling of the relevant probability distribution. While straightforward in principle, difficulties can arise in practice if the probability distribution involved is "sparse". In the case of sparse distributions the diffusive random walk methods generally used to perform the sampling can become problematic as transitions from one isolated region of importance to another grow rare on a practical scale. Unfortunately, such rare-event sampling difficulties are themselves not rare events. They arise frequently, for example, in the treatment of thermally activated processes.[6]

To focus the present discussion, we consider the problem of estimating an average of the form

$$\langle V \rangle = \frac{\int \mu(\mathbf{x}) V(\mathbf{x}) d\mathbf{x}}{\int \mu(\mathbf{x}) d\mathbf{x}},$$

(1.1)

where $\mathbf{x}$ represents the coordinates of the system and $V(\mathbf{x})$ is a property of interest such as the potential energy. As long as the probability distribution involved, $\mu(\mathbf{x})$, has a single "inherent structure"[7,8], the sampling is straightforward and the associated numerical results are reliable. If, on the other hand, the probability distribution has multiple, isolated inherent structures, this ceases to be the case. In such cases $<V>$ can be thought of as a sum of contributions over the various inherent structures,

$$\langle V \rangle = \sum_\alpha \langle V \rangle_\alpha W_\alpha.$$

(1.2)

In Eq. (1.2) $<V>_\alpha$ represents the average of V over the region corresponding to inherent structure $\alpha$, $\mathfrak{R}_\alpha$

$$\langle V \rangle_\alpha = \frac{\int_{\mathfrak{R}_\alpha} \mu(\mathbf{x}) V(\mathbf{x}) d\mathbf{x}}{\int_{\mathfrak{R}_\alpha} \mu(\mathbf{x}) d\mathbf{x}},$$

(1.3)

and $W_\alpha$ is the fraction of the statistical weight associated with that region

$$W_\alpha = \frac{\int_{\mathfrak{R}_\alpha} \mu(\mathbf{x}) d\mathbf{x}}{\int \mu(\mathbf{x}) d\mathbf{x}}.$$

(1.4)

Even though one does not typically decompose the required averages into explicit component form, as emphasized by Eq. (1.2) both intra and inter-inherent structure sampling are at issue in their calculation. The relevant point is that while conventional sampling methods do well for the former task, their performance for the latter can be unreliable.

Methods designed to deal with rare-event sampling issues have been developed and are briefly summarized in Section II. While many represent a significant improvement over basic methods, the need for more effective techniques remains an important area of current research. In the present work we describe a new approach that is based on the use of symmetrization. At first glance, the conscious introduction of this element into the discussion is counter-intuitive in that its presence is generally regarded as increasing rather than decreasing the complexity of the problem. As shown in the following sections, however, because it alters the inherent connectedness of the probability distributions involved, symmetrization represents a potentially useful tool in the treatment of rare-event problems.

The outline of the remainder of the paper is as follows: Section II presents both a brief overview of current approaches to the rare-event sampling problem and a heuristic discussion of the present technique. For reasons that will be discussed in greater detail in Section II, we denote this method the "infinite swapping" or INS technique. The details of the INS approach and practical methods for its implementation are then outlined in Section III. Numerical applications that compare the performance of INS and parallel tempering methods for the calculation of thermodynamic properties of selected Lennard-Jones (LJ) clusters are discussed in Section IV. Finally, Section V summarizes our findings and indicates possible directions for future development.

## II. Background and Observations:

A number of methods for dealing with rare-event sampling problems have been developed and are discussed elsewhere.[2,9-13] Common strategies include changes in trial moves, changes of measure and parallel tempering[10,11] or replica exchange methods.[12]

Perhaps the simplest device for dealing with rare-event problems is to increase the length scale(s) associated with trial displacements. Although such increases generally result in diminished acceptance probabilities, they can nonetheless sometimes improve the ability of the associated random walks to overcome barriers. If one has sufficient physical insight concerning the nature of the relevant potential energy surfaces involved, "displacement vector Monte Carlo" techniques[14] become an option and can be effective.

A general technique for dealing with rare-event issues is through a change of measure.[2,3] For example, one can rewrite the average in Eq. (1.1) as

$$\langle V \rangle = \frac{\left\langle \left( \frac{\mu(\mathbf{x})}{\tilde{\mu}(\mathbf{x})} \right) V(\mathbf{x}) \right\rangle_{\tilde{\mu}}}{\left\langle \left( \frac{\mu(\mathbf{x})}{\tilde{\mu}(\mathbf{x})} \right) \right\rangle_{\tilde{\mu}}},$$

(2.1)

where the averages on the right hand side of Eq. (2.1) are over an auxillary probability distribution, $\tilde{\mu}(\mathbf{x})$, that can be chosen for sampling convenience. When adopting this approach one must guard against $\tilde{\mu}(\mathbf{x})$ choices that cause unacceptable increases in the associated variance. A number of strategies for $\tilde{\mu}(\mathbf{x})$ selection have been proposed.[2,3,15,16]

A widely utilized method for dealing with rare-event sampling issues is the parallel tempering[10,11,13] or replica exchange[12] technique. This approach utilizes an expanded computational ensemble composed of systems corresponding to different values of one or more of the control variables. In the study of activated processes, for example, temperature is a natural parameter. Parallel tempering tries to overcome rare-event sampling issues by exchanging information between different portions of the simulation. Information from higher temperatures, where high-energy barriers are more easily crossed, is used to improve the sampling at lower temperatures, where such crossings are more difficult. Strategies for the selection of the tempering ensemble have been discussed.[17-21]

Information transfer within the computational ensemble is essential to the parallel tempering approach. It is important to note that the nature of the exchange moves that are typically used to generate this transfer places practical limits on the rate of information flow achievable within the approach. For example, conventional, single-temperature coordinate displacements are typically augmented with trial moves that involve attempted swaps of system coordinates *between* temperatures. The percentage of swap attempts used is generally set using the Goldilocks[22] - "not-too-small/not-too-big" - approach. If the percentage is too small, the information flow between temperatures is minimal and the resulting sampling benefits are negligible. If, on the other hand, the percentage is too large, the sampling again suffers. In the extreme limit, where *all* trial moves become attempted swaps, fixed system configurations are ineffectively toggled back and forth between the various temperature streams.

Using the theory of large deviations, it is possible to show that the performance of parallel tempering is, in principle, a monotonically increasing function of the swap rate.[23] Empirical indications of such behavior have been noted.[24] Were it possible to design such a procedure, this result suggests that the infinite swapping limit of parallel tempering would have desirable performance characteristics. Our analysis shows that it is, in fact, possible to construct an alternative scheme, which coincides with the infinite swapping limit of parallel tempering in a distributional sense and which can be readily implemented. More generally, the method that emerges is a special case of a new class of rare-event sampling techniques.

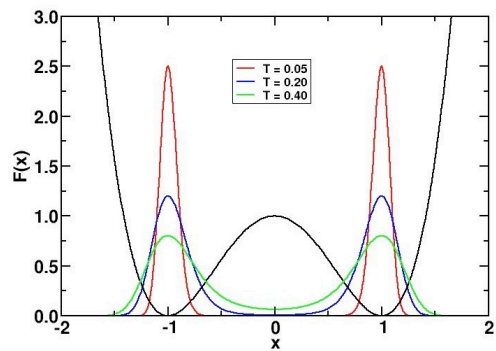

Figure 1: Plots of V(x) for the double-well example (Eq. (2.2)) and three normalized Boltzmann distributions (Eq. (2.3)) corresponding to T= 0.05, 0.20, 0.40.

We illustrate the basic ideas that emerge from our large deviations analysis by considering a simple example, a double well system whose potential energy is given by

$$V(x) = (x^2 - 1)^2.$$
(2.2)

In Fig. (1) we present plots of the potential energy and of the classical thermal distributions of the form

$$\pi(x,T) = e^{-V(x)/T},$$
(2.3)

for three temperatures, here chosen to be T = 0.05, 0.20, and 0.40. As expected, the normalized probability densities shown in Fig. (1) are increasingly "sparse" at lower temperatures where the scale of thermal fluctuations becomes small relative to the energy barrier that separates the wells. In Fig. (2) we present the information of Fig. (1) in a somewhat different manner. Specifically, we define a multi-variable distribution, $\mu_0$, as

$$\mu_0(x,y,z,T_x,T_y,T_z) = \pi(x,T_x)\pi(y,T_y)\pi(z,T_z), \quad (2.4)$$

and assign the three temperatures used in Fig. (1) to particular coordinates, ($T_x$=0.05, $T_y$=0.20, $T_z$=0.40). The extent to which the various potential minima are linked is reflected in the structure of the isosurfaces of $\mu_0$ shown in Fig. (2). The greatest bridging density is associated with the z-direction (highest temperature) and the least with the x-direction (lowest temperature). In Fig. (3) we present an isosurface plot of a related distribution, $\mu_{xyz}$, the fully symmetrized analog of the original. This (unnormalized) distribution is defined as

$$\begin{aligned}\mu_{xyz}(x,y,z,T_x,T_y,T_z) = &\ \mu_0(x,y,z,T_x,T_y,T_z) + \mu_0(x,y,z,T_x,T_z,T_y) \\ &+ \mu_0(x,y,z,T_y,T_x,T_z) + \mu_0(x,y,z,T_y,T_z,T_x) \\ &+ \mu_0(x,y,z,T_z,T_x,T_y) + \mu_0(x,y,z,T_z,T_y,T_x).\end{aligned}$$

(2.5)

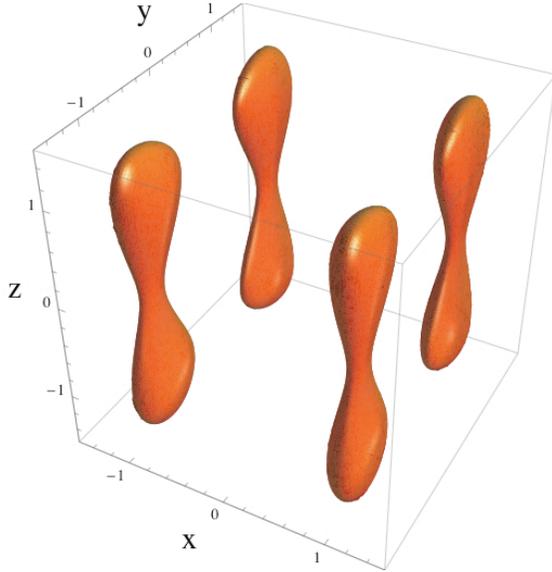

Figure 2: Plot of the isosurface $\mu_0(x,y,z) = 0.05$, where $\mu_0$ is defined in Eq. (2.4).

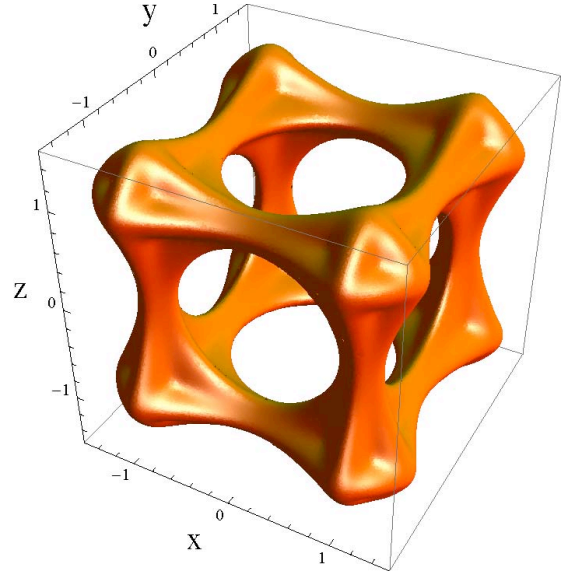

Figure 3: Plot of the isosurface $\mu_{xyz}(x,y,z) = 0.05$, where $\mu_{xyz}$ is defined in Eq. (2.5).

Because symmetrization removes the explicit linkage between temperature and coordinate labels, $\mu_{xyz}$ is more highly connected than $\mu_0$. Thus, while both distributions contain related thermodynamic information, Figs. (2) and (3) suggest that the symmetrized version offers fewer rare-event sampling issues than the original. Whether or not this idea can be exploited and the factorial-scale growth of the computational complexity associated with total symmetrization can be avoided remains to be demonstrated.

One possible strategy is suggested in Fig. (4). These plots display isosurfaces of two, partially symmetrized distributions, $\mu_{xy}$ and $\mu_{yz}$, defined for the present double-well example as

$$\mu_{xy}(x,y,z) = \mu_0(x,y,z,0.20,0.40,0.05) + \mu_0(x,y,z,0.40,0.20,0.05) \quad (2.6)$$

and

$$\mu_{yz}(x,y,z) = \mu_0(x,y,z,0.05,0.20,0.40) + \mu_0(x,y,z,0.05,0.40,0.20). \quad (2.7)$$

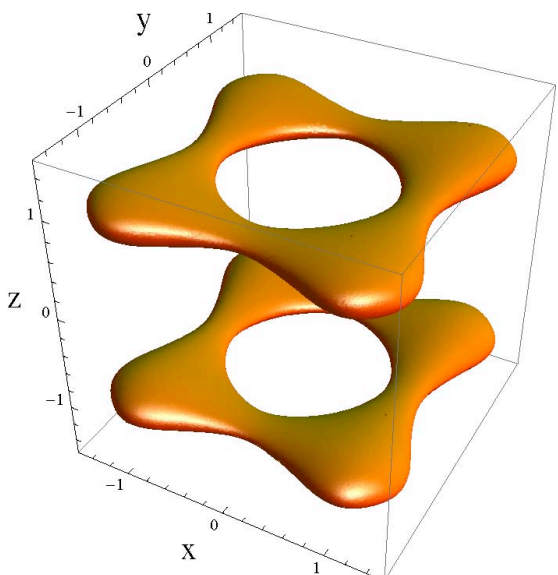
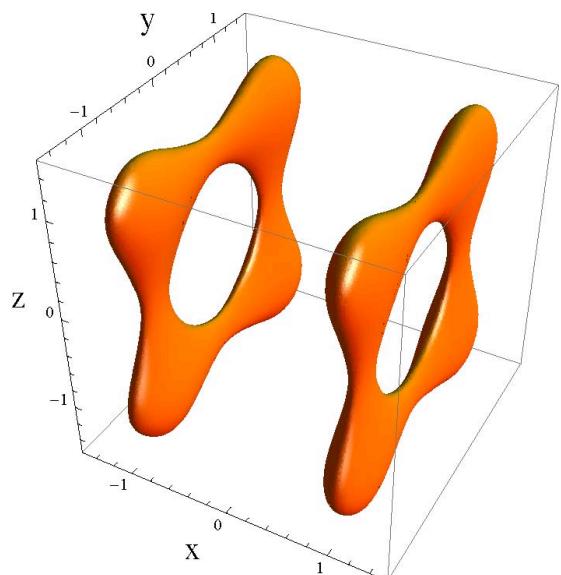

Figure 4: (a) Plot of the isosurface $\mu_{xy}(x,y,z) = 0.05$, where $\mu_{xy}$ is defined in Eq. (2.6), (b) Plot of the isosurface $\mu_{yz}(x,y,z) = 0.05$, where $\mu_{yz}$ is defined in Eq. (2.7).

Figure 4: (b) Plot of the isosurface $\mu_{yz}(x,y,z) = 0.05$, where $\mu_{yz}$ is defined in Eq. (2.7).

Although not as great as that seen for full symmetrization, we see in Fig. (4) that partial symmetrization produces an increase in the connectedness of the associated distribution, a suggestion that the approach may offer a practical means of achieving the sampling benefits we seek.

Before discussing the details of the approach, it is useful to consider symmetrization from a somewhat different perspective. Equation (2.4) represents a typical parallel tempering ensemble; a simple product of Boltzmann terms each corresponding to a specified temperature. The mixing of coordinate and temperature information that underlies the method's improved sampling is generated indirectly, through the choice of trial moves. The structure of the symmetrized distribution defined in Eq. (2.5), on the other hand, intrinsically entangles temperature and coordinate information. All coordinate displacements, even those that would take place in a single temperature setting for the distribution in Eq. (2.4), thus occur in a Born-Oppenheimer like environment generated by the (infinitely rapid) multi-temperature swaps of information inherent in the symmetrization process.

***III. Methods:*** In this section we describe the INS approach and develop methods for its practical implementation. The formal basis of the method is discussed elsewhere.[23] We present the approach in the context of a representative equilibrium problem, that of estimating the average potential energy of a system characterized by a set of coordinates, **x**, a known potential energy, $V(\mathbf{x})$, and a specified probability distribution, $\pi(\mathbf{x}, T)$. We assume that our objective is to compute the average potential energy at a specified temperature (or temperatures), $T_k$,

$$<V>_k = \frac{\int \pi(\mathbf{x},T_k)V(\mathbf{x})d\mathbf{x}}{\int \pi(\mathbf{x},T_k)d\mathbf{x}}.$$

(3.1)

The joint probability distribution for N independent systems with coordinates, $\{\mathbf{x}_n\}$ and a set of temperatures, $\{T_n\}$, n=1,N (assumed to include $T_k$) is given by the product $\pi(\mathbf{x}_1,T_1)\pi(\mathbf{x}_2,T_2)...\pi(\mathbf{x}_N,T_N)$ and its associated partition function is given by

$$Z = \prod_{n=1}^{N}\int d\mathbf{x}_n \pi(\mathbf{x}_n,T_n).$$

(3.2)

In what follows we denote the set of permutations of the temperatures between the coordinate sets by $\{P_n[\pi(\mathbf{x}_1,T_1)\pi(\mathbf{x}_2,T_2)...\pi(\mathbf{x}_N,T_N)]\}$, where n=1,N! and the coordinate set label corresponding to the temperature $T_k$ in the $n^{th}$ permutation by the function indx(n,k). In terms of these quantities, we can write $<V>_k$ as the average of N! formally identical terms

$$<V>_k = \frac{1}{N!}\sum_n \frac{\int P_n[\pi(\mathbf{x}_1,T_1)\pi(\mathbf{x}_2,T_2)..\pi(\mathbf{x}_N,T_N)]V(\mathbf{x}_{indx(n,k)})d\mathbf{X}}{\int P_n[\pi(\mathbf{x}_1,T_1)\pi(\mathbf{x}_2,T_2)..\pi(\mathbf{x}_N,T_N)]d\mathbf{X}},$$

(3.3)

where **X** represents the N sets of system coordinates, $(\mathbf{x}_1,\mathbf{x}_2,....,\mathbf{x}_N)$. Because all of the normalization integrals for the terms in Eq. (3.3) are identical, the average involved can also be written as

$$<V>_k = \frac{\int \mu(\mathbf{X})\left[\sum_n \rho_n(\mathbf{X})V(\mathbf{x}_{indx(n,k)})\right]d\mathbf{X}}{\int \mu(\mathbf{X})d\mathbf{X}},$$

(3.4)

where

$$\rho_n(\mathbf{X}) = \frac{P_n[\pi(\mathbf{x}_1,T_1)\pi(\mathbf{x}_2,T_2)..\pi(\mathbf{x}_N,T_N)]}{\mu(\mathbf{X})},$$

(3.5)

and where the (fully) thermally symmetrized distribution, $\mu(\mathbf{X})$, is given by

$$\mu(\mathbf{X}) = \sum_n P_n[\pi(\mathbf{x}_1,T_1)\pi(\mathbf{x}_2,T_2)..\pi(\mathbf{x}_N,T_N)].$$

(3.6)

The sum of the statistical weights defined in Eq. (3.5) is, by construction, equal to unity. For a specified temperature, the averages defined in Eqs. (3.1) and (3.4) are formally identical. However, as discussed in Section II, the increased connectedness of $\mu(\mathbf{x})$ suggests that $\mu(\mathbf{x})$ is a useful importance function and that Eq. (3.4) offers a potential computational advantage over Eq. (3.1) with respect to rare-event issues.

Equations (3.4)-(3.6) represent the full, N-temperature infinite swapping method. The invariant distribution involved is the sum of N! terms arising from all possible temperature permutations. If the number of temperatures is small, it is possible to sample this distribution directly. Even though complete INS sampling will become impractical beyond a modest number of temperatures, its consideration is useful since it provides both insights and benchmark examples that will prove useful for the development and validation of more generally applicable "partial" INS (PINS) approaches.

The fully symmetrized distribution is well defined and can be sampled using a variety of Monte Carlo techniques. We outline one approach in Appendix A and illustrate its application with a simple example. We find this approach valuable since, with suitable extension, it is well suited for both full and partial INS applications.

Simple card games played an important role in the discovery of modern Monte Carlo methods.[25] They also provide a useful framework for understanding the partial swapping approach we describe in Appendix B. Generating all possible permutations of the symmetrized distribution $\mu(\mathbf{x})$ is analogous to the problem of achieving all possible orderings in a deck of cards. Card shuffling techniques can be designed in a variety of ways. The most straightforward is to shuffle the entire deck at once. Alternatively, one can perform a multi-player shuffle in which players individually partition the deck into subsets, shuffle the cards within each of the subsets generated by their partitioning and then pass the deck to another player. As long as the partitionings involved do not share a common boundary, the resulting process will produce a complete shuffling. A simple, two-player, three-card example serves to illustrate the basic idea. Assume that when they have control of the deck, the players' actions are as follows:
- Player-1: leave top card in place, shuffle cards two and three
- Player-2: shuffle top two cards, leave card three in place.

Although neither player's individual actions would do so, the actions of the two players combined will shuffle the entire deck. Randomization is produced locally by the individual players and globally by their interaction. It is important to note that, even when the number of cards (temperatures) involved is large, the explicit randomization steps involve subsets whose sizes can be made controllably small. We outline a general "dual" (two-player) sampling procedure based on this idea in Appendix B.

It is useful to note that the methods described in Appendix A and B contain internal diagnostics that can be helpful in numerical applications. Specifically, the statistical weights associated with the various permutations, $\rho_n(\mathbf{x})$, contain information that provides insight into the nature of the sampling. If only a small number of the permutations in the relevant tempering ensemble have significant statistical weight, the sampling benefits associated with symmetrization will be minimal. This situation might arise, for example, if the spacings between the temperatures in the ensemble are too large. If, on the other hand, the ensemble temperatures are well chosen, then the statistical weight will be spread more broadly across the computational ensemble. The Shannon entropy associated with the $\rho$-values involved,

$$S_\rho = -\sum_n \rho_n \log(\rho_n),$$

(3,7)

is a measure of the dispersion of such $\rho$-values and thus provides a useful device for monitoring the sampling.

*IV. Numerical Applications:* We now present a series of studies designed to explore the utility of the infinite and partial infinite-swapping approaches. Because they provide a common environment for few and many-body systems, exhibit a range of computational challenges, and have an extensive published literature, we have chosen Lennard-Jones clusters as the subject for these investigations.[26] We begin with a series of relatively simple applications that illustrate the basic elements of the present approach and that provide a convenient means for comparing the performance of infinite swapping methods to that of current techniques. After these initial tests, we present a series of applications to systems of increasingly demanding rare-event character in order to provide a basis for gauging the performance of the new techniques.

===================================================================

*Table I:*

Shown are the average values of the potential energy (in units of $\varepsilon$) for a four-atom LJ system for five different reduced temperatures ($k_BT/\varepsilon$). Results are computed by a variety of methods, including full infinite swapping (INS), various sub-infinite approaches that are discussed more fully in the text, and by conventional Metropolis techniques. The numbers quoted in parantheses are the standard deviation estimates for the last digit quoted in the associated results.

| $k_BT/\varepsilon$ | INS | 1-4/4-1 | 2-3/3-2 | 1-2-2/2-2-1 | Metropolis |
|---|---|---|---|---|---|
| 0.0500 | -5.8370 (3) | -5.8374 (3) | -5.8375 (3) | -5.8378 (4) | -5.8366 (7) |
| 0.0600 | -5.8007 (3) | -5.8012 (4) | -5.8010 (4) | -5.8013 (4) | -5.8017 (9) |
| 0.0800 | -5.7219 (5) | -5.7223 (5) | -5.7227 (6) | -5.7225 (6) | -5.7225 (12) |
| 0.1000 | -5.6300 (8) | -5.6305 (10) | -5.6310 (9) | -5.6297 (10) | -5.6319 (17) |
| 0.1200 | -5.5191 (18) | -5.5184 (20) | -5.5184 (18) | -5.5190 (20) | -5.5178 (24) |

===================================================================

Unless otherwise stated, the following computational details apply to all studies discussed in the present section. Interaction potentials are assumed to be a sum of pairwise Lennard-Jones interactions of the form,

$$v(r) = 4\varepsilon\left(\left(\frac{\sigma}{r}\right)^{12} - \left(\frac{\sigma}{r}\right)^{6}\right), \quad (4.1)$$

where $\varepsilon$ and $\sigma$ are energy and length scale parameters, respectively and r is the inter-particle separation. Dimensionless energy and length scales, set by $\varepsilon$ and $\sigma$, are used throughout. Temperatures quoted are in dimensionless form ($k_BT/\varepsilon$) and all times are expressed in natural Lennard-Jones units, $(\varepsilon/m\sigma^2)^{1/2}$, where m is the atomic mass. Cluster evaporation is controlled by means of the customary addition of a confining potential to the pairwise Lennard-Jones interactions.[27] The form of the confining potential used in the present study is

$$V_c(\mathbf{r_1},\mathbf{r_2},...,\mathbf{r}_N) = \sum_{i=1}^{N} v_c(\mathbf{r}_i), \quad (4.2)$$

where

$$v_c(\mathbf{r}) = \varepsilon\left(\frac{|\mathbf{r}-\mathbf{r}_{cm}|}{R_c}\right)^{20}, \quad (4.3)$$

and where $\mathbf{r}_{cm}$ and $R_c$ are the cluster center of mass and constraining radius, respectively. In order to provide a uniform basis for comparison of the various sampling methods being discussed, we utilize a common technique, the smart Monte Carlo (SMC) method,[28] to perform the necessary sampling moves. Our choice of the SMC approach is based on a variety of considerations. At the practical level, SMC methods make use of familiar and readily available molecular dynamics techniques, thus making them relatively easy to implement. In addition, dynamically based SMC methods are well-suited for treating the correlated, many-particle barrier dynamics involved in the activated processes we are investigating. In the SMC approach "moves" are generated by assigning Gaussian random momenta characteristic of the temperature in question to all Cartesian degrees of freedom for the cluster and then propagating the classical equations of motion forward for a specified length of time via molecular dynamics methods. As defined, SMC "moves" thus involve the displacement of all cluster coordinates. Subsequent moves are generated by repeating this process with the last configuration serving as the starting point for the next move. Provided that the total system energy is conserved for the molecular dynamics segments, the configurations so produced provide a proper sampling of the Boltzmann distribution of positions at the specified temperature. Because the time scale for the molecular-dynamics segments involved is relatively short (on the order of the cluster vibrations), it is relatively easy to achieve an acceptable level of energy conservation. Fourth-order Runge-Kutta methods with step sizes of between 0.005 and 0.010 Lennard-Jones time units are utilized for all molecular dynamics integration tasks in the present work.[29]

We present in Table I a series of average potential energies computed for the simple, four-atom LJ cluster at five temperatures. These results are computed for the various temperatures using a number of different approaches, including the ordinary Metropolis technique, INS, and three different PINS approaches. To facilitate comparison, all results are generated using $2 \times 10^5$ SMC moves of 0.50 time duration and a confining potential parameter of $R_c = 2.5$. Of these moves, $5 \times 10^4$ are used as warm-up. Because there is only one stable potential minimum, rare-event sampling issues are not a concern for this system. What is relevant is that we see in Table I that all methods yield consistent estimates of the average potential energy at the various temperatures involved. In particular, the average potential energies produced by the full INS method, various types of PINS sampling, and single-temperature Metropolis sampling all agree.

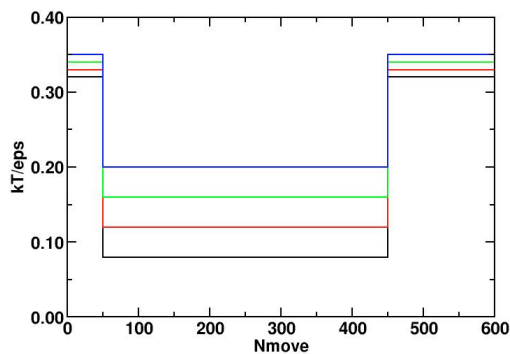

Figure 5: (a) Temperature profiles for the four-temperature, LJ-13 relaxation simulations. $N_{move}$ refers to the number of single-temperature Monte Carlo moves in the thermal cycle.

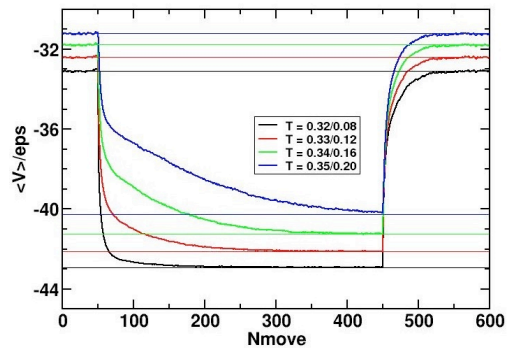

Figure 5: (b) <V> values for the LJ-13 infinite swapping relaxation simulations performed using the temperature profiles in Fig. (5a).

We now wish to compare the performance of a number of different sampling methods for a series of increasingly demanding applications. "Relaxation experiments," of the type illustrated in Fig. (5) for a simple four-temperature 13-atom LJ cluster, are a convenient device for making such comparisons. The idea behind such numerical experiments is to impose a periodic temperature variation in the computational ensemble and to monitor the response of a calculated average property, in this case the average potential energy. One period of the temperature variations used in the present four-temperature, 13-atom LJ study is shown in Fig. (5a). In this example one period consists of 600 SMC moves, with moves 50-450 being the "cooling" portion and the remainder the "heating" portion of the cycle. As this cycle is repeated, potential energies corresponding to various numbers of SMC moves are accumulated and "relaxation" curves of the type shown in Fig. (5b) are generated. The results in Fig. (5b) are generated using 2000 heating/cooling cycles of the type displayed in Fig. (5a) using complete INS sampling. SMC moves of 0.50 duration and a confining potential parameter of $R_c = 2.5$ are used. For comparison purposes, the values of the average potential energies for the high and low-temperatures shown in Fig. (5b) are computed in separate INS simulations using $6 \times 10^5$ SMC moves. The rates at which the heating/cooling curves approach their limiting equilibrium values can be seen in Fig. (5b).

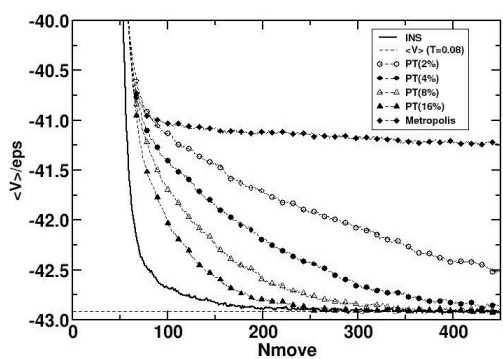

Figure 6: The cooling portion of the of the lowest of the four-temperature studies of Fig. (5b). Results correspond to infinite swapping (INS), parallel tempering (PT), and ordinary Metropolis sampling. The dashed line corresponds to the numerically exact value of <V> for T = 0.08.

Relaxation studies allow us to compare the rates of equilibration for different sampling methods. In Fig. (6), for example, we compare the cooling portions of the lowest of the four temperatures shown in Fig. (5) obtained using various sampling methods, including INS sampling, parallel tempering, and conventional single-temperature Metropolis methods. The parallel tempering results shown in Fig. (6) are labeled with the percentages of trial swap moves involved. Specifically, the percentage stated corresponds to the probability that during each update of the coordinates there is an attempted parallel tempering exchange between one, randomly chosen, nearest-neighbor *pair* of temperatures. In this notation, conventional, single-temperature Metropolis results correspond to the zero swapping limit of parallel tempering. The other computational parameters used for the studies of Fig. (6) are the same as those in Fig. (5). To provide a common basis for comparison, all studies in Fig. (6), including the conventional Metropolis approach, utilize SMC sampling moves of duration 0.50. We see that the observed rates of equilibration for the sampling methods vary significantly, with conventional, single-temperature Metropolis sampling being the slowest and the INS approach the fastest. We see in Fig. (6) that over the range typically utilized in parallel tempering simulations the rate of equilibration is an increasing function of the percentage of pair swap attempts.[24] As discussed in Section II and as illustrated in Fig. (7), the improvement in the convergence rate (as measured by the value of <V> at a particular point on the relaxation curve) ultimately saturates and reverses as the percentage of attempted pair swaps is increased. We note in passing that the "optimal" swap percentage for this example is larger than that commonly utilized in tempering applications.

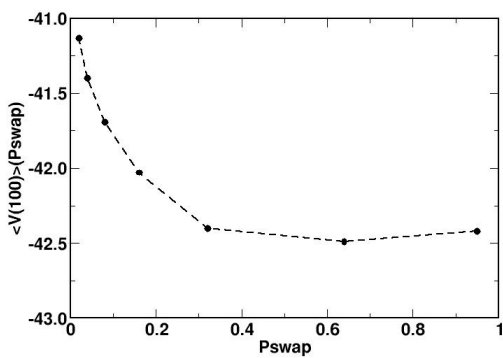

Figure 7: A plot of a representative value on the parallel tempering relaxation curves of Fig. (6) ($N_{move} = 100$) as a function of the attempted pair swap probability ($P_{swap}$).

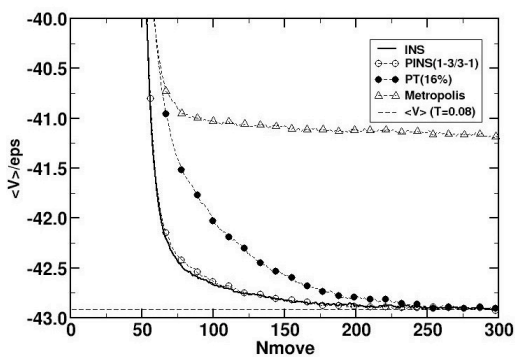

Figure 8: As in Fig. (6) with the addition of the partial swapping results (PINS).

We conclude our LJ-13 study by examining the quality of the partial infinite swapping approach. In Fig. (8) we compare the cooling curves for the INS and parallel tempering results of Fig. (6) with that obtained using a dual-chain PINS approach. Using the notation of Appendix B, both PINS chains are composed of two temperature blocks, a lowest temperature and a group of three higher temperatures for one chain and group of three lower temperatures and a single highest temperature for the other. We see that the performance of this PINS(1-3/3-1) approach is quite similar to that of the full infinite-swapping method. For example, the value of <V> for the PINS(1-3/3-1) cooling curve at $N_{move} = 100$, the point used in Fig. (7) to gauge the rate of parallel tempering equilibration, is -42.64, quite close to that produced by the full INS simulation (-42.66) and better than the "optimal" parallel tempering result of Fig. (7) (-42.49). Although encouraging, conclusions concerning the performance and general utility of the PINS method await more serious tests, tests we now consider.

The 38-atom Lennard-Jones cluster is an appreciably more complex system than those we have considered thus far. As discussed elsewhere,[30] the potential energy surface for this system exhibits a multi-funnel structure in which the global and lowest-lying local minimum differ in energy by an amount that is small relative to the barrier that separates them. This and the presence of roughly $4 \times 10^{14}$ local minima[31] are indications that the LJ-38 system presents non-trivial computational challenges.

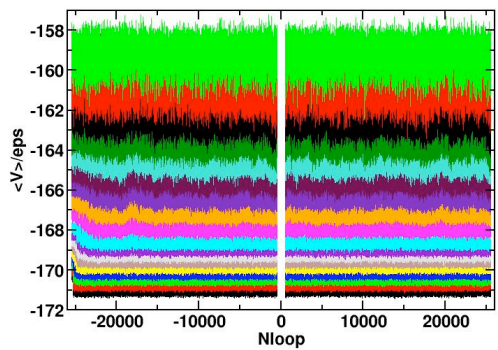

Figure 9: Loop averages (100 moves/loop) of the potential energy of a 38-atom LJ cluster. Results are obtained using a 6-temperature floating PINS simulation starting from a "melt" configuration (left edge to center) and from a second simulation starting from the known global minimum geometry (right edge to center). The bands correspond to the potential energies of 18 temperatures from the range (0.05,0.18).

Although we will ultimately adopt other, more effective implementations, we begin our LJ-38 investigations with a series of applications that utilize a rather basic version of the PINS approach, the "floating" partial swapping method. As discussed in Appendix B, this simple approach utilizes a symmetrized band of temperatures that moves randomly within the larger computational ensemble. Using these methods, we examine the issue of equilibration in the LJ-38 system. In Fig. (9) we compare the loop averages (100 SMC moves per loop) for two different six-temperature floating PINS simulations. Both simulations use ensembles that contain 45-temperatures that span the range from (0.050-0.210) in steps of 0.005 and the range from (0.210 - 0.330) in steps of 0.010. SMC moves of 1.00 duration and a confining potential parameter of $R_c = 2.65$ are used. Loop averages shown in Fig. (9) correspond to the potential energies for 18 of the 45 temperatures and cover the range from 0.050-0.180. Results shown on the left in Fig. (9) are from a simulation that is initiated using a randomly chosen, high-temperature "melt" configuration (a down-anneal) while those on the right are from a second simulation that is initiated using the known global minimum configuration (an up-anneal). For ease of visual comparison of the final results, the output for the up-anneal simulation is displayed in reverse in Fig. (9). That is, the results of the two simulations begin at the edges and end in the center of the figure. We see that the loop averages generated in the up and down anneals for the various temperatures are consistent (i.e. merge as they approach the center of the figure). As further evidence of this consistency, the heat capacities computed using the last 20,000 loops or $2 \times 10^6$ SMC moves of both simulations are compared in Fig. (10). The heat capacities for these two simulations, including the "shoulder" region between temperatures of 0.10 and 0.15, are in good agreement.

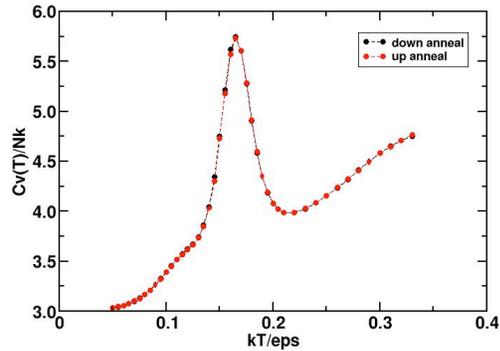

Figure 10: Heat capacities for the LJ-38 system. The "down" ("up") labels refer to the corresponding simulations of Fig. (9).

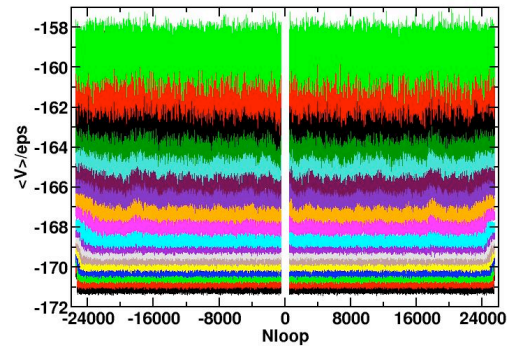

Figure 11: As in Fig. (9), except that the right hand portion of the plot corresponds to results obtained starting with the icosahedral isomer instead of the global minimum energy structure.

In Figs. (11) and (12) we present additional LJ-38 equilibration and heat capacity studies. In these figures the up-anneal results of Figs. (9) and (10) are replaced with analogous six-temperature floating PINS results in which the initial configuration is taken to be that of the lowest-lying LJ-38 icosahedral isomer instead of the global minimum. Significant equilibration difficulties have been reported for parallel tempering simulations initialized using this particular isomer.[32] We see that the quality of the PINS results in Figs. (11) and (12) are similar to those shown in Figs. (9) and (10), a further indication that PINS methods are effective in coping with the rare-event sampling issues presented by the LJ-38 system.

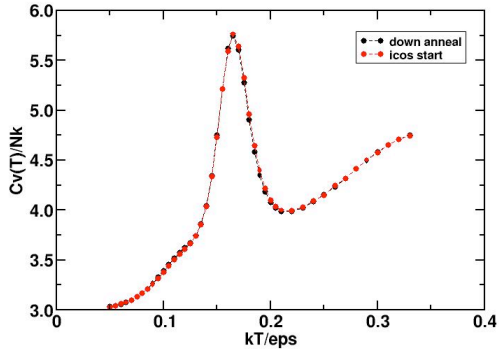

Figure 12: Heat capacities for the LJ-38 system extracted from the last 20,000 loops of the simulations of Fig (11).

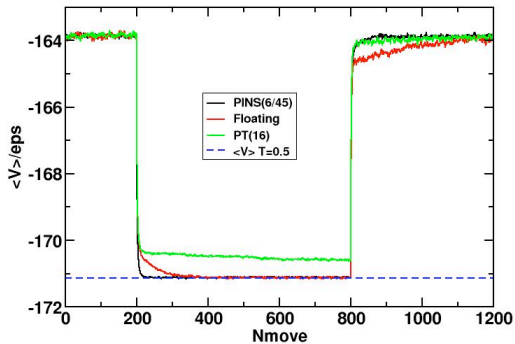

Figure 13: Relaxation simulations for LJ-38. The <V> values shown are for the lowest temperature of the 45-temperature ensemble (see text for details) and are obtained using parallel tempering (PT), a "floating" six-temperature partial swapping approach, and dual-chain PINS(6/45) methods. The numerically exact results for <V> = 0.05 are shown for comparison.

The LJ-38 results shown in Figs. (9-12) indicate that the floating PINS approach is a useful technique. It is important to note, however, that other, more effective PINS methods are available. As an illustration we present in Fig. (13) the results of a series of numerical relaxation experiments for the LJ-38 system. These are obtained using parallel tempering (16% swap rate), floating PINS and dual-chain PINS approaches for the 45 temperature ensemble described previously. For simplicity, only results for the lowest temperature in the computational ensemble are shown. Because our floating PINS simulation is based on a six-temperature symmetrized block, we also use a dual-chain PINS study in which sampling chains are composed of the same size blocks. Specifically, each of the chains is composed of seven, six-temperature blocks plus one smaller block of three temperatures for a total of 45 temperatures. In one chain, the three-temperature block is made up of the lowest three temperatures, in the other chain the highest three. Heating and cooling profiles used are of the generic type illustrated in Fig. (5a). In the present study, however, the cycles consist of 1200 SMC moves, each of one unit Lennard-Jones time duration. The cooling segment is taken as the portion of the cycle from moves 200 to 800 with the remainder being the heating portion. During the cooling portion of the cycle the 45 temperatures in the ensemble cover the range from (0.050-0.210) in temperature steps of 0.005, and from (0.210 - 0.330) in steps of 0.010 while during the heating portion of the cycle temperatures less than or equal to 0.150 are set equal to 0.150. It should be noted that for a fixed block size, the computational effort in the dual-chain PINS approach grows linearly in the total number of temperatures. All simulations utilize a constraining radius of $R_c = 2.65$ and are obtained using 600 thermal cycles. We see in Fig. (13) that equilibration is achieved by both the floating and dual-chain PINS approaches and that the rates for both are significantly better than those of parallel tempering. In fact, we see evidence of incomplete equilibration in the parallel tempering results of Fig. (13). We also see that the rates of equilibration for the dual-chain PINS approach are significantly better than those for the simple floating PINS approach. It is important to note that this improvement in equilibration rates is achieved with a relatively modest increase in computational effort. For example, the computing times required for the floating and dual-chain PINS results in Fig. (13) are only 9% and 12% more than those required for the corresponding parallel tempering results, respectively.

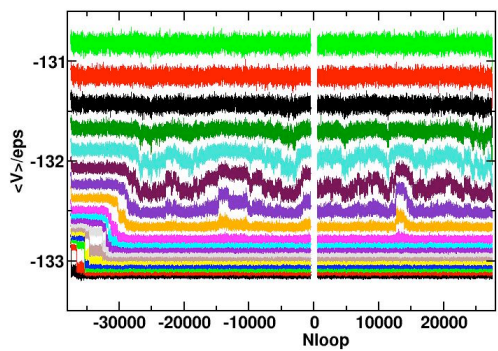

Figure 14: PINS(4/54) equilibration studies for the LJ-31 system. Temperatures in the ensemble are distributed geometrically in the range (0.01,0.35). Loop averages for 18 temperatures in the range (0.01,0.0535) are shown. Results on the left (right) side of the plot are initialized using a high-temperature melt configuration (the known global minimum geometry).

Finally, we consider the 31-atom Lennard-Jones cluster, a system that offers a number of interesting features. As discussed by Doye, et al.,[30] the potential energy surface for this system is relatively complex with many, nearly degenerate local minima separated by large barriers. This complex energy landscape makes locating the global minimum structure difficult[30] and produces a low-temperature heat capacity feature that represents a challenging computational objective.[33] In Fig. (14) we present down and up-anneal studies for the LJ-31 system. The results shown are obtained using two PINS(4/54) simulations. As with the previous LJ-38 studies, the down-anneal starts from a randomly chosen, high-temperature "melt" configuration while the up-anneal is initiated using the known global minimum. The 54 temperatures in the ensemble are distributed geometrically in the range (0.01,0.35). Loop averages for 18 temperatures over the range (0.01,0.0535) are shown in Fig. (14). These simulations contain various numbers of loops that consist of 100 SMC moves per loop, with each SMC move being of unit time duration. The confining radius used in these LJ-31 studies is $R_c = 3.00$, the same as that in Ref. (33). The down-anneal simulation uses 37,000 total loops with the final 25,000 being used for data collection. Because the warm-up period involved is appreciably smaller, the up-anneal uses 27,000 total loops, again with the final 25,000 being for data collection.

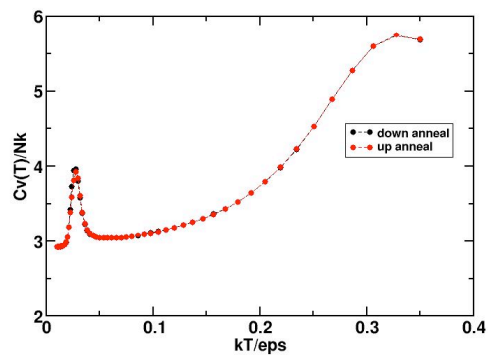

Figure 15: The heat capacities of the LJ-31 system as a function of temperature extracted from the final 25,000 loops of the PINS(4/54) simulations of Fig. (14).

The heat capacities for the LJ-31 system computed from these two simulations are shown in Fig. (15). As was the case with the analogous LJ-38 studies, the agreement of the heat capacities produced by the up and down-anneal simulations is quite good, including the region of the low-temperature heat capacity peak near T = 0.027. The potential energy fluctuations associated with rare-event hops between the various local minima that produce this low-temperature heat capacity feature are evident in Fig. (14) in the potential energy range of roughly -132.5 ± 0.5.

*V. Discussion and Summary:* The rare-event sampling problem is a troublesome and ubiquitous one. In the present work we have developed an approach for dealing with such problems that is based on the use of symmetrization. In addition, we have presented the results for an initial series of applications to a representative class of problems, the calculation of the thermodynamic properties of Lennard-Jones clusters. Our preliminary findings suggest that the infinite swapping approach is an effective tool for dealing with rare-event problems and represents an improvement over parallel tempering.

It is important to note that, while both the infinite swapping and parallel tempering approaches utilize an expanded computational ensemble to improve their rare-event sampling performance, there is an important distinction between the two approaches. In parallel tempering the invariant distribution is a simple product of individual Boltzmann components. The transfer of temperature-related information used to improve the rare-event sampling is achieved by expanding the space of possible trial moves to include swaps of data between the various temperature streams. In contrast, the invariant distribution for the infinite swapping approach is a thermally symmetrized sum of Boltzmann-like terms. This distribution is more highly connected than the original, unsymmetrized form. Moreover, its basic structure intrinsically entangles temperature and coordinate information thereby increasing the flow of temperature-related information within the computational ensemble. This increased flow of information is at the core of the infinite swapping approach. Practical methods for sampling the associated distributions have been presented and discussed.

Beyond the classical statistical-mechanical applications considered in the present work, infinite swapping methods would appear to have other areas of potential utility. One obvious area is that of minimization.[29] Preliminary applications suggest that the increased flow of temperature-related information makes INS-inspired approaches useful for such applications and that further investigations are warranted. Quantum simulations, where the problems of sampling symmetrized[34] and sparse distributions[35] arise naturally in path-integral simulations, are other areas of possible payoff.

*Acknowledgments:* The authors gratefully acknowledge grant support of this research through the DOE Multiscale and Optimization for Complex Systems program No. DE-SC0002413. NP wishes to thank the Swiss National Science Foundation for postdoctoral support and JDD wishes to acknowledge support through DOE departmental program No. DE-00015561. PD wishes to acknowledge support from the Army Research Office (W911NF-09-1-0155) and PD, HW and YL gratefully acknowledge support from the National Science Foundation (DMS-1008331). The authors would also like to thank Professors D. L. Freeman and H. Jónsson and Drs. K. Spiliopoulos and Cristian Predescu for valuable discussions concerning the present work.

*Appendix A:*

We describe the infinite swapping sampling approach by illustrating its application to the calculation of a representative thermodynamic property, the average potential energy at a specified temperature, $T_k$. For simplicity, we will assume that this temperature is one of the set of ensemble temperatures $\{T_j\}$, j=1,N. The N sets of system coordinates, $(\mathbf{x}_1,\mathbf{x}_2,....,\mathbf{x}_N)$, are collectively denoted as $\mathbf{X}$. The fully symmetrized distribution is defined in Eq. (3.6) and the statistical weight, $\rho_n(\mathbf{x}_1,\mathbf{x}_2,....,\mathbf{x}_N)$, for each of the N! possible permutations is given by Eq. (3.5). Assuming that the configuration at the m$^{th}$ step is denoted by $(\mathbf{x}_{1,m},\mathbf{x}_{2,m},....,\mathbf{x}_{N,m})$, or as simply $\mathbf{X}_m$, the sampling process includes the following steps:

- calculate the ρ-weights for the current configuration, $\{\rho_n(\mathbf{X}_m)\}$, for all permutations in the N-temperature ensemble, n=1,N!
- select one of the ρ-weights randomly according to its size
- generate a new configuration, $\mathbf{X}_{m+1}$, by making single-temperature moves for each of the coordinate sets using the temperature-coordinate associations inherent in the randomly chosen permutation
- calculate the ρ-weights for all permutations in the N-temperature ensemble for the new configuration, $\{\rho_n(\mathbf{X}_{m+1})\}$, n=1,N!
- update the accumulating sum being used to assemble the $<V>_k$ average by adding

$$\Delta_{m+1}(k) = \sum_{j=1}^{N} V(\mathbf{x}_{j,m+1})\Gamma(j,k)$$ to the k$^{th}$ sum, where $\Gamma(j,k)$ is equal to the sum of the ρ-weights for all permutations in which coordinate set j is associated with temperature $T_k$.

- repeat the process using the new configuration, $\mathbf{X}_{m+1}$, as input

As an illustration, we follow the above process through a hypothetical cycle for a one-dimensional system and a three-temperature ensemble. If we denote the permutation in which $x_K$ is associated with temperature 1, $x_L$ with temperature 2, etc. by [K,L,...], then the six possible permutations arising from our three-temperature ensemble are given by

| n | $P_n$ |
|---|---|
| 1 | [1,2,3] |
| 2 | [1,3,2] |
| 3 | [2,1,3] |
| 4 | [2,3,1] |
| 5 | [3,1,2] |
| 6 | [3,2,1] |

Given a particular configuration, $(x_{1,m},x_{2,m},x_{3,m})$ we first evaluate the ρ-values associated with each of the possible permutations and then select one of these permutations randomly in proportion to its size. For purposes of illustration, we assume that permutation 3 in the above table is the permutation selected. Single-temperature Monte Carlo moves would then be made in which coordinate 2 would be advanced using temperature $T_1$, coordinate set 1 using temperature $T_2$, and coordinate 3 using temperature $T_3$. The potential energies, $\{V(x_j)\}$, and ρ-weights, $\{\rho_n(x_j)\}$, j=1,3, n=1,6 corresponding to the new coordinates would be computed and the three accumulating sums being used in the calculation of the potential energy averages would be updated as

$$\text{Sum}(T_1) \rightarrow \text{Sum}(T_1) + V(x_1)(\rho_1 + \rho_2) + V(x_2)(\rho_3 + \rho_4) + V(x_3)(\rho_5 + \rho_6)$$
$$\text{Sum}(T_2) \rightarrow \text{Sum}(T_2) + V(x_1)(\rho_3 + \rho_5) + V(x_2)(\rho_1 + \rho_6) + V(x_3)(\rho_2 + \rho_4)$$
$$\text{Sum}(T_3) \rightarrow \text{Sum}(T_3) + V(x_1)(\rho_4 + \rho_6) + V(x_2)(\rho_2 + \rho_5) + V(x_3)(\rho_1 + \rho_3).$$

## *Appendix B:*

In this appendix we outline the partial INS sampling approach. As in Appendix A, we assume that we are dealing with an N-temperature ensemble and that our objective is the calculation of thermodynamic properties such as the average potential energy. The methods we outline are designed for the calculation of such single-temperature properties. With suitable extension more general properties, for example those that depend on the joint distribution of multiple temperatures, can also be treated.[23]

The procedure we describe is a "dual-chain" process in which the full, N-temperature set is partitioned into blocks in two distinct ways. Unlike the full INS approach, symmetrization is partial and is confined to the various temperature blocks that make up the two chains. We label the two chains as α and β and denote the temperatures for block-1 of chain-α as $\{T_j^{\alpha_1}\}$, those for block-2 of the chain-β as $\{T_j^{\beta_2}\}$, etc. The system configuration at the m$^{th}$ step in the sampling is denoted as $(x_{1,m}, x_{2,m}, ..., x_{N,m})$ or as simply $\mathbf{X}_m$. When necessary, partition and temperature block labels can also be included. For example, the coordinates at the m$^{th}$ sampling step of block-1 of chain-α are labeled as $\mathbf{X}_m^{\alpha_1}$, for block-2 of chain-β as $\mathbf{X}_m^{\beta_2}$, and so on. Unless stated otherwise we assume that the partitionings of chains α and β remain fixed during the simulation.

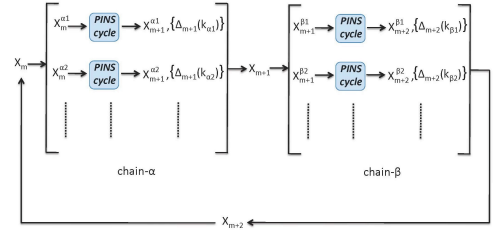

Figure 16: A schematic depiction of the PINS sampling approach utilized in the present studies.

The overall structure of the dual-chain PINS sampling method is depicted schematically in Fig. (16). Local mixing within blocks is produced by symmetrization while larger-scale mixing is generated by the exchange of information between chains. The individual chains are consistent with multiple invariant distributions, namely those that have the proper relative weights of the permutations within the various blocks. The unique distribution that the chains share is the fully symmetrized distribution. Consequently, while neither chain will do so individually, the two chains working in tandem will, if suitably designed, generate a proper sampling of the fully symmetrized distribution, Eq. (3.6).[23] It is important to note that the PINS approach avoids dealing with all possible permutations simultaneously. Because the two chains utilize different partitionings, a proper sampling requires that we specify rules both to govern the construction of the required thermodynamic averages from the information produced by the individual chains and to control the "hand-off" of data between them.  In our applications thermodynamic information emerges in the form of update increments that are accumulated to produce the average potential energies at the various temperatures.  In Fig. (16) the potential energy increments for the temperatures in the first block of chain-α at step m+1 are denoted as $\Delta_{m+1}(k_{\alpha_1})$, those at step m+2 for the temperatures in the second block of chain-β as $\Delta_{m+2}(k_{\beta_2})$, and so on.

We first describe the PINS sampling process in general terms and then illustrate its use by considering a particular example.  Assuming that the m+1$^{st}$ step starts with temperature block-1 of chain-α, the detailed steps in the "PINS cycle" depicted in Fig. (16) are as follows:

- define $\mathbf{Y}_m = \mathbf{X}_m^{\alpha 1}$
- calculate the ρ-weights, $\{\rho_n(\mathbf{Y}_m)\}$, n=1,N($\alpha_1$), for all N($\alpha_1$) permutations in block $\alpha_1$
- select one of the ρ-weights randomly in proportion to its size
- generate a new configuration, $\mathbf{Y}_{m+1}$, using the temperature-coordinate associations inherent in the randomly chosen permutation
- calculate the new ρ-weights, $\{\rho_n(\mathbf{Y}_{m+1})\}$, n=1,N($\alpha_1$), for all N($\alpha_1$) permutations in block $\alpha_1$
- calculate the potential energy update increments for the temperatures in the block
$$\Delta_{m+1}(k) = \sum_{j=1}^{N(\alpha_1)} V(\mathbf{y}_{j,m+1})\Gamma(j,k),$$
where $\Gamma(j,k)$ is equal to the sum of the ρ-weights for all permutations of block $\alpha_1$ in which coordinate set j is associated with temperature $T_k$.
- select one of the newly generated ρ-weights randomly in proportion to its size
- use the temperature-coordinate associations in the randomly selected permutation to generate the coordinates for block $\alpha_1$ at the m+1$^{st}$ step, $\mathbf{X}_{m+1}^{\alpha_1}$, to be handed off to the complementary chain

These steps are then repeated for the remaining blocks in chain-α.  After all chain-α steps are complete, the $\mathbf{X}_{m+1}^{\alpha}$ coordinates for the various blocks are merged to form the new configuration, $\mathbf{X}_{m+1}$, and this configuration is passed on to chain-β as indicated in Fig. (16).  Chain-β then produces a new configuration, $\mathbf{X}_{m+2}$, which is then passed back to chain-α, etc. and the process is repeated as necessary.

We illustrate the PINS procedure by considering its application to the three-temperature example of Appendix A using a dual-chain PINS(1-2/2-1) approach.  This designation indicates that chain-α consists of a single-temperature block at temperature $T_1$ and a two-temperature, symmetrized block involving the temperatures $T_2$ and $T_3$.  In chain-β, on the other hand, the two-temperature block is made up of temperatures $T_1$ and $T_2$ while the single temperature block involves only $T_3$.  This system is the analog of the three-card example discussed near the end of Section III.  For simplicity, we assume that the configuration at the m$^{th}$ sampling step is given by $\mathbf{X}_m = (x_{1,m}, x_{2,m}, x_{3,m})$ and that the next sampling move involves chain-α.  Defining $\mathbf{Y}_m = \mathbf{X}_m$, the blocks and the ρ-values for the permutations in the two blocks in chain-α are evaluated numerically as:

| Block | n | $\rho_n(\mathbf{Y}_m)$ |
|---|---|---|
| 1 | 1 | 1 |
| 2 | 1 | $\pi(y_{2,m},T_2)\pi(y_{3,m},T_3)/(\pi(y_{2,m},T_2)\pi(y_{3,m},T_3) + \pi(y_{3,m},T_2)\pi(y_{2,m},T_3))$ |
|   | 2 | $\pi(y_{3,m},T_2)\pi(y_{2,m},T_3)/(\pi(y_{2,m},T_2)\pi(y_{3,m},T_3) + \pi(y_{3,m},T_2)\pi(y_{2,m},T_3))$, |

where $\pi(y,T)$ is the single-temperature Boltzmann distribution defined by Eq. (2.3). In block-1 of chain-$\alpha$, the coordinate-temperature association involves $y_1$ and $T_1$. Consequently, $y_{1,m+1}$ is generated from $y_{1,m}$ using the temperature $T_1$ and any suitable, single-temperature Monte Carlo procedure. Smart Monte Carlo methods[28] are used for the applications in Section IV. Since there is only one coordinate and temperature for block-1, $x_{1,m+1} = y_{1,m+1}$ and the potential energy increment is $\Delta_{m+1}(T_1) = V(y_{1,m+1})$. To generate the sampling moves for block-2 of chain-$\alpha$, we evaluate the block's $\rho$-values, $\rho_1$ and $\rho_2$, for the variables $(y_{2,m},y_{3,m})$ and select one of the $\rho$-values randomly in proportion to its size. For illustration purposes we assume that $\rho_2$ is selected. We then make a Monte Carlo move from $y_{2,m}$ to $y_{2,m+1}$ and $y_{3,m}$ to $y_{3,m+1}$ using the temperature-coordinate associations in the selected term. From the last line of the table above, we see that the temperature-coordinate associations in $\rho_2$ involve the pairings $y_2$-$T_3$ and $y_3$-$T_2$. Consequently, we advance $y_2$ at a temperature $T_3$ and $y_3$ at a temperature $T_2$. Had $\rho_1$ been chosen instead, the $y_2$ advance would have been made using the temperature $T_2$ and the corresponding $y_3$ advance using the temperature $T_3$. Using the new configuration $\mathbf{Y}_{m+1} = (y_{2,m+1},y_{3,m+1})$, we re-evaluate the block-2 $\rho$-values, $\rho_1(\mathbf{Y}_{m+1})$ and $\rho_2(\mathbf{Y}_{m+1})$. These newly generated $\rho$-values are then used to produce the potential energy increments for $T_2$ and $T_3$ as well as the values of $(x_{2,m+1},x_{3,m+1})$ that will be "handed-off" to chain-$\beta$. The potential energy increments are given by

$$\Delta_{m+1}(T_2) = V(y_{2,m+1})\,\rho_1(\mathbf{Y}_{m+1}) + V(y_{3,m+1})\,\rho_2(\mathbf{Y}_{m+1})$$
$$\Delta_{m+1}(T_3) = V(y_{2,m+1})\,\rho_2(\mathbf{Y}_{m+1}) + V(y_{3,m+1})\,\rho_1(\mathbf{Y}_{m+1})$$

To generate the values of $(x_{2,m+1},x_{3,m+1})$ that, along with $x_{1,m+1}$, will be passed on to chain-$\beta$, we select one of the new $\rho$-values randomly in proportion to its size and assign the $(x_{2,m+1},x_{3,m+1})$ values based on the coordinate-temperature associations in the selected term. The results for the two possible selections are

| term selected | y-T associations | x-assignments |
|---|---|---|
| $\rho_1(\mathbf{Y}_{m+1})$ | $y_2$-$T_2$, $y_3$-$T_3$ | $x_{2,m+1} = y_{2,m+1}$ |
|   |   | $x_{3,m+1} = y_{3,m+1}$ |
| $\rho_2(\mathbf{Y}_{m+1})$ | $y_2$-$T_3$, $y_3$-$T_2$ | $x_{2,m+1} = y_{3,m+1}$ |
|   |   | $x_{3,m+1} = y_{2,m+1}$ |

The final x-values so generated, $\mathbf{X}_{m+1} = (x_{1,m+1},x_{2,m+1},x_{3,m+1})$, are then handed off as input to chain-β and an analogous series of steps are made using that chain's partitioning. The $\mathbf{X}_{m+2}$ values produced by the β-chain are then handed back into the α-chain and the entire dual process is repeated.

We have found it convenient to work with partitionings that are composed of multiple blocks that contain the same, even number of temperatures in combination with a single block that contains half that number of temperatures. To assure that they share no common temperature boundaries, the smaller block in one chain is made up of the lowest temperatures of the ensemble while in the other it involves the highest. The major block size and total number of temperatures thus serve as identifiers of the sampling chains used in the simulation. In this more compressed notation, for example, our previous PINS(1-2/2-1) illustration is designated as a PINS(2/3) simulation.

One has significant flexibility in designing many of the details of the PINS sampling approach. Decisions concerning the potential merits of many of these various options await further study. One option is the choice of the single-temperature sampling method. Although all numerical examples presented here use smart Monte Carlo methods,[28] any technique that provides a proper sampling of the Boltzmann distributions involved can be utilized for the single-temperature sampling tasks. We note that while the infinite sampling approach is guaranteed to perform better than parallel tempering, the degree of improvement can depend on the particular sampling used for the single-temperature moves. Another example is the possible use of multi-chain methods. It is straightforward to formulate multi-chain sampling methods that are generalizations of the dual-chain techniques utilized in the present studies. Whether or not these, either by themselves or in combination with particular partitioning strategies, offer a computational advantage is an interesting and open question. Another example is the use of multi-step methods. In the present work only a single sampling step is made within each of the chains before handing-off the calculation to the complementary chain. It is important to note that multiple steps are permitted, a fact that may prove of practical significance in the context of parallel implementations of the PINS approach. Finally, we note that the partitionings of the various chains need not remain fixed during the simulation. In the present work, for example, the "floating" PINS applications are implemented using a dynamical version of dual-chain sampling in which the chain partitions vary as a single, symmetrized block of temperatures moves randomly throughout the larger computational ensemble.